 \documentclass[aps,pra,twocolumn,groupedaddress,amsmath,amssymb,showpacs]{revtex4}
\usepackage{graphicx}
\usepackage{amsmath}
\usepackage{dcolumn}% Align table columns on decimal point
\usepackage{bm}% bold math

\begin{document}
\title{Maximally entangled mixed states made easy}
\author{A. Aiello}
\author{G. Puentes}
\author{D. Voigt}
\author{J.P. Woerdman}
\affiliation{Huygens Laboratory, Leiden University\\
P.O.\ Box 9504, 2300 RA Leiden, The Netherlands}
\begin{abstract}
We show that it is possible to achieve maximally entangled mixed
states of two qubits from the singlet state via the action of
local non-trace-preserving quantum channels. Moreover, we present
a simple, feasible linear optical implementation of one of such
channels.
\end{abstract}

 \maketitle
%
%03.65.Ud    Entanglement and quantum nonlocality (e.g. EPR
%paradox, Bell's inequalities, GHZ states, etc.) (for entanglement
%production in quantum information, see 03.67.Mn; for entanglement
%in Bose-Einstein condensates, see 03.75.Gg)
%
%03.65.Nk    Scattering theory
%
%03.67.Mn    Entanglement production, characterization, and
%manipulation (see also 03.65.Ud Entanglement and quantum
%nonlocality; for entanglement in Bose-Einstein condensates, see
%03.75.Gg)
%
%42.25.Ja    Polarization
%
%42.50.Dv    Nonclassical states of the electromagnetic field,
%including entangled photon states; quantum state engineering and
%measurements (see also 03.65.Ud Entanglement and quantum
%nonlocality, e.g. EPR paradox, Bell's inequalities, GHZ states,
%etc.)
%
%
\section{introduction}
In a recent paper, Ziman and Bu\v{z}ek have demonstrated that it
is impossible to transform the singlet state of two qubits in a
maximally entangled mixed state (MEMS), via local  maps
$\mathcal{E}\otimes \mathcal{I}$ \cite{Ziman1}. Such maps describe
the action of quantum channels $\mathcal{C}_\mathcal{E}$ acting on
a single qubit of the initial singlet state. When a channel is
local, that is when it acts on a single qubit, the corresponding
map is subjected to some restrictions. This can be easily
understood in the following way: Let Alice and Bob be two
spatially separated observer who can make measurements on qubits
$a$ and $b$, respectively, and let $\rho_\mathrm{in}$ and
$\rho_\mathrm{out}$ denote the density matrices describing the
two-qubit quantum state before and after the channel,
respectively. In absence of any causal connection between Alice
and Bob, special relativity demands that Bob cannot detect via any
type of local measurement the presence of the channel
$\mathcal{C}_\mathcal{E}$ in the path of photon $a$. Since the
physics of qubit $b$ is described  to Bob by the reduced density
matrix $\rho_\mathrm{out}^\mathrm{B}  = \mathrm{Tr}
\rho_\mathrm{out} |_a$, the locality constraint can be written as
\begin{equation}\label{i10}
\rho_\mathrm{out}^\mathrm{B} = \rho_\mathrm{in}^\mathrm{B}.
\end{equation}
If we write explicitly the map $\mathcal{E} \otimes \mathcal{I}$
as
\begin{equation}\label{i20}
\rho_\mathrm{in} \mapsto \rho_\mathrm{out} =\sum_\mu \left(A_\mu
\otimes I \right) \rho_\mathrm{in} \left(A_\mu^\dagger \otimes I
\right),
\end{equation}
then Eq. (\ref{i10}) becomes
\begin{equation}\label{i30}
\sum_{k,l}(\rho_\mathrm{in})_{li,kj} \Bigl( \sum_\mu A_\mu^\dagger
A_\mu \Bigr)_{kl} = \sum_{k} ( \rho_\mathrm{in} )_{ki,kj},
\end{equation}
which implies the \emph{trace-preserving} condition on the local
map $\mathcal{E} \otimes \mathcal{I}$:
\begin{equation}\label{i40}
 \sum_\mu
A_\mu^\dagger A_\mu  = I.
\end{equation}
Local maps that do not satisfy Eq. (\ref{i40}) are classified as
\emph{non-physical}, and are not investigated in Ref.
\cite{Ziman1}.

In this paper we show that under certain circumstances, it may be
meaningful to consider the action of non-trace-preserving maps, as
well. In particular, we give two simple examples of local
non-trace-preserving maps that generate maximally entangled mixed
states of two qubits from the singlet state. Two-qubit MEMS states
may exist in two subclasses usually denoted as MEMS I and MEMS II
\cite{Munro1}. In Section II,  we furnish an explicit
representation for two maps $\mathcal{M}$ and $\mathcal{K}$ that
generate MEMS I and II states, respectively. In Sec. III a
feasible linear optical implementation of the quantum channel
$\mathcal{C}_\mathcal{M}$ corresponding to the map $\mathcal{M}$
is given. In Sec. IV we introduce an all-unitary linear optical
model for  $\mathcal{C}_\mathcal{M}$ and, via a rigorous QED
treatment, we show how the ``non-physical'' map $\mathcal{M}$
arises in a natural manner. Finally, we draw our conclusions in
Sec. V.
\section{Non-trace-preserving maps}
In this section we introduce two non-trace-preserving maps
$\mathcal{M}$ and $\mathcal{K}$ that generate MEMS I and II
states, respectively, from an initial singlet state of two qubits.
\subsection{MEMS I map $\mathcal{M}$}
Let $\rho_\mathrm{in}$ represent the initial state of a single
qubit that is transformed under the action of the map
$\mathcal{M}$ as: $\rho_\mathrm{in} \mapsto \rho_\mathrm{out}$,
where
\begin{equation}\label{a10}
\rho_\mathrm{out} = \sum_{\mu = 0}^3 \mathbf{M}_\mu
\rho_\mathrm{in} \mathbf{M}_\mu^\dagger,
\end{equation}
and $\mathbf{M}_0 = \mathbf{0} = \mathbf{M}_1$,
\begin{equation}\label{a20}
\mathbf{M}_2 = \sqrt{2(1-p)} \left(%
\begin{array}{cc}
  1 & 0 \\
  0 & 0 \\
\end{array}%
\right), \qquad \mathbf{M}_3 = \sqrt{p} \left(%
\begin{array}{cc}
  0 & -1 \\
  1 & 0 \\
\end{array}%
\right),
\end{equation}
where $2/3 \leq p \leq 1$. This map is \emph{not} trace-preserving
nor unital, since
\begin{equation}\label{a30}
\sum_{\mu = 0}^3 \mathbf{M}_\mu^\dagger \mathbf{M}_\mu = \sum_{\mu
= 0}^3 \mathbf{M}_\mu \mathbf{M}_\mu^\dagger =
\left(%
\begin{array}{cc}
  2 - p & 0 \\
  0 & p \\
\end{array}%
\right) \neq \mathbf{I}_2,
\end{equation}
where  $\mathbf{I}_n$ denotes the $n \times n$ identity matrix.
The apparent non-physical nature of this map can be displayed if
we use the Pauli matrices to rewrite
\begin{equation}\label{a40}
\mathbf{M}_2 = \sqrt{\frac{1-p}{2}} (\mathbf{I}_2 + { \bm \sigma
}_z), \qquad \mathbf{M}_3 = - i \sqrt{p} \, { \bm \sigma }_y,
\end{equation}
and  substitute Eq. (\ref{a40}) into Eq. (\ref{a10}) to obtain
\begin{equation}\label{a50}
\begin{array}{rcl}
\rho_\mathrm{out} & = & \displaystyle{\frac{1}{2} \bigl[(1 -
p)\rho_\mathrm{in} + 2 p \, { \bm \sigma }_y \rho_\mathrm{in} {
\bm \sigma }_y + (1-p)\, { \bm \sigma }_z \rho_\mathrm{in}
{ \bm \sigma }_z \bigr.} \\\\
&& \bigl. + (1-p) \left\{\rho_\mathrm{in}, { \bm \sigma }_z
\right\} \bigr],
\end{array}
\end{equation}
where the anti-commutator term ($\left\{ a,b \right\} = ab + ba$)
is clearly responsible for non conservation of the trace.

Now, let us consider the map $\mathcal{M}$ as representative of
the local quantum channel $\mathcal{C}_\mathcal{M}$ acting on a
single qubit belonging to an entangled pair prepared in the
initial state $\rho_\mathrm{in}$ (note that \emph{now}
$\rho_\mathrm{in}$ denotes a two-qubit state, therefore it is
represented by a $4 \times 4$ matrix). The two-qubit map
$\mathcal{M}$ can be written as
\begin{equation}\label{a60}
\rho_\mathrm{out} = \sum_{\mu = 0}^3 \mathbf{N}_\mu
\rho_\mathrm{in} \mathbf{N}_\mu^\dagger,
\end{equation}
where $\mathbf{N}_\mu =\mathbf{M}_\mu \otimes \mathbf{I}_2$,
namely  $\mathbf{N}_0 = \mathbf{0} = \mathbf{N}_1$, and
\begin{subequations}
\label{a70}
\begin{equation}
\begin{array}{rcl}
\mathbf{N}_2 & = & \sqrt{2(1-p)} \left(%
\begin{array}{cccc}
  1 & 0 & 0 & 0 \\
  0 & 1 & 0 & 0 \\
  0 & 0 & 0 & 0 \\
  0 & 0 & 0 & 0 \\
\end{array}%
\right),
\end{array}
\end{equation}
\begin{equation}
\begin{array}{rcl}
 \mathbf{N}_3 & = & \sqrt{p} \left(%
\begin{array}{cccc}
  0 & 0 & -1 & 0 \\
  0 & 0 & 0 & -1 \\
  1 & 0 & 0 & 0 \\
  0 & 1 & 0 & 0 \\
\end{array}%
\right).
\end{array}
\end{equation}
\end{subequations}
Let $| \phi^- \rangle = (|01 \rangle - |10 \rangle)/\sqrt{2}$ be
the two-qubit input singlet state represented by the density
matrix $\rho_\mathrm{s}$:
\begin{equation}\label{a80}
\rho_\mathrm{s} = | \phi^- \rangle \langle \phi^- | = \left(%
\begin{array}{cccc}
  0 & 0 & 0 & 0 \\
  0 & \frac{1}{2} &  -\frac{1}{2} & 0 \\
  0 &  -\frac{1}{2} &  \frac{1}{2} & 0 \\
  0 & 0 & 0 & 0 \\
\end{array}%
\right).
\end{equation}
A straightforward calculation shows that
\begin{equation}\label{a90}
\rho_\mathrm{I} =  \sum_{\mu = 0}^3 \mathbf{N}_\mu \rho_\mathrm{s}
\mathbf{N}_\mu^\dagger =  \left(%
\begin{array}{cccc}
  \frac{p}{2} & 0 & 0 & \frac{p}{2} \\
  0 & 1-p &  0& 0 \\
  0 &  0 &  0 & 0 \\
  \frac{p}{2} & 0 & 0 & \frac{p}{2} \\
\end{array}%
\right)
\end{equation}
which represent a  MEMS I state.
\subsection{MEMS II map $\mathcal{K}$}
As before, let $\rho_\mathrm{in}$ represent the initial state of a
single qubit that transforms under the action of the map
$\mathcal{K}$ as: $\rho_\mathrm{in} \mapsto \rho_\mathrm{out}$,
where
\begin{equation}\label{a230}
\rho_\mathrm{out} = \sum_{\mu = 0}^3 \mathbf{K}_\mu
\rho_\mathrm{in} \mathbf{K}_\mu^\dagger,
\end{equation}
where $\mathbf{K}_1 = \mathbf{0}$, and
\begin{subequations}\label{a240}
\begin{equation}
\mathbf{K}_0  =  \displaystyle{  \sqrt{\frac{2}{3}} \left(%
\begin{array}{cc}
  1 & 0 \\
  0 & 0 \\
\end{array}%
\right)} ,
\end{equation}
\begin{equation}
 \mathbf{K}_2  =  \displaystyle{ \sqrt{\frac{1}{3} - \frac{p}{2}} \left(%
\begin{array}{cc}
  0 & 1 \\
  1 & 0 \\
\end{array}%
\right),}
\end{equation}
\begin{equation}
 \mathbf{K}_3  =  \displaystyle{ \sqrt{\frac{1}{3} + \frac{p}{2}} \left(%
\begin{array}{cc}
  0 & -1 \\
  1 & 0 \\
\end{array}%
\right)},
\end{equation}
\end{subequations}
and $0\leq p \leq 2/3$. As in the case of $\mathcal{M}$, this map
is not trace-preserving nor unital, since
\begin{equation}\label{a250}
\sum_{\mu = 0}^3 \mathbf{K}_\mu^\dagger \mathbf{K}_\mu =\sum_{\mu
= 0}^3 \mathbf{K}_\mu \mathbf{K}_\mu^\dagger = \frac{2}{3}
\left(%
\begin{array}{cc}
  2  & 0 \\
  0 & 1 \\
\end{array}%
\right) \neq \mathbf{I}_2.
\end{equation}
A straightforward calculation shows that the \emph{two-qubit} map
$\mathcal{K}$ realized by $\mathbf{L}_\mu = \mathbf{K}_\mu \otimes
\mathbf{I}_2$ produces MEMS II states when acting upon the singlet
state (\ref{a80}):
\begin{equation}\label{a260}
\rho_\mathrm{II} =  \sum_{\mu = 0}^3 \mathbf{L}_\mu
\rho_\mathrm{s}
\mathbf{L}_\mu^\dagger =  \left(%
\begin{array}{cccc}
  \frac{1}{3} & 0 & 0 & \frac{p}{2} \\
  0 & \frac{1}{3} &  0& 0 \\
  0 &  0 &  0 & 0 \\
  \frac{p}{2} & 0 & 0 & \frac{1}{3} \\
\end{array}%
\right),
\end{equation}
where $\mathbf{L}_1 = \mathbf{0}$, and
\begin{subequations}\label{a270}
\begin{equation}
\mathbf{L}_0  =  \displaystyle{  \sqrt{\frac{2}{3}} \left(%
\begin{array}{cccc}
  1 & 0 & 0 & 0 \\
  0 & 1 & 0 & 0 \\
  0 & 0 & 0 & 0 \\
  0 & 0 & 0 & 0 \\
\end{array}%
\right)} ,
\end{equation}

\begin{equation}
 \mathbf{L}_2  =  \displaystyle{ \sqrt{\frac{1}{3} - \frac{p}{2}} \left(%
\begin{array}{cccc}
  0 & 0 & 1 & 0 \\
  0 & 0 & 0 & 1 \\
  1 & 0 & 0 & 0 \\
  0 & 1 & 0 & 0 \\
\end{array}%
\right),}
\end{equation}

\begin{equation}
 \mathbf{L}_3  =  \displaystyle{ \sqrt{\frac{1}{3} + \frac{p}{2}} \left(%
\begin{array}{cccc}
  0 & 0 & -1 & 0 \\
  0 & 0 & 0 & -1 \\
  1 & 0 & 0 & 0 \\
  0 & 1 & 0 & 0 \\
\end{array}%
\right)}.
\end{equation}
\end{subequations}
\section{Linear optical implementation of the channel $\mathcal{C}_\mathcal{M}$}
\begin{figure}[!htl]
\includegraphics[angle=0,width=7.5truecm]{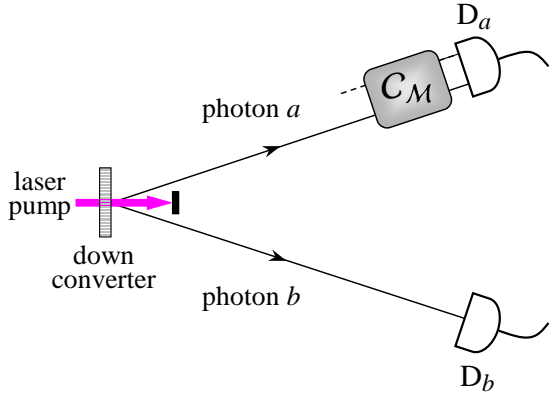}
\caption{\label{fig:A}  Sketch of the proposed experimental setup.
The box $\mathcal{C}_\mathcal{M}$ represents the  quantum channel.
A photon from an intense laser pump is split into the pair $(a,b)$
by the down-converter. Detectors $\mathrm{D}_a$ and $\mathrm{D}_b$
permit a tomographically complete reconstruction of the two-photon
quantum state. Further details are given in the text.}
\end{figure}

\begin{figure}[!hbr]
\includegraphics[angle=0,width=8truecm]{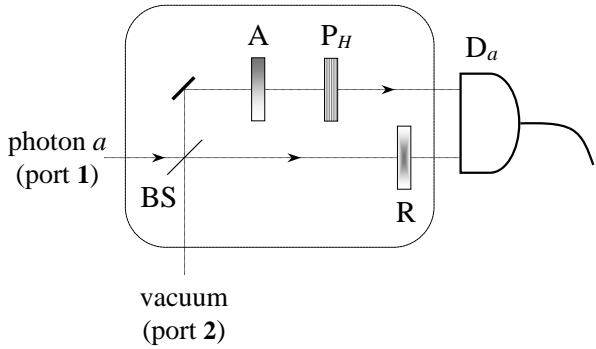}
\caption{\label{fig:B} Detailed scheme of the quantum channel
$\mathcal{C}_\mathcal{M}$. BS denotes a $50/50$ beam splitter,
$\mathrm{A}$ is a beam attenuator, $\mathrm{P}_H$ is a linear
polarizer that selects horizontally polarized photons, and
$\mathrm{R}$ is a
 polarization rotator oriented at $\theta = \pi /2$. }
\end{figure}

The layout of the experiment we propose to create MEMS I states is
illustrated schematically in Fig. \ref{fig:A}.
Two photons in the singlet state $(|HV\rangle - |VH
\rangle)/\sqrt{2}$ emerge from the down-converter. Here $H$ and $V
$ are labels for horizontally and vertically polarized photons,
respectively. Photon $b$ goes directly to detector $\mathrm{D}_b$,
while photon $a$ goes to $\mathcal{C}_\mathcal{M}$ and then to
detector $\mathrm{D}_a$. $\mathcal{C}_\mathcal{M}$ is a linear
optical two-port device that is illustrated in detail in Fig.
\ref{fig:B}. Supposedly, detector $\mathrm{D}_a$ does not
distinguish which output port of $\mathcal{C}_\mathcal{M}$ the
photon comes from: This is our mechanism to induce decoherence.
Photon $a$ enters port $\mathbf{1}$ and can be either transmitted
to path 1 or reflected to path 2 by the $50/50$ beam splitter BS;
vacuum
enters port $\mathbf{2}$. Let the square bracket vectors $\left[%
\begin{array}{c}
  1 \\
  0 \\
\end{array}%
\right]$ and $\left[%
\begin{array}{c}
  0 \\
  1 \\
\end{array}%
\right]$ represent the two orthogonal \emph{position} states (or,
spatial modes) of a photon travelling in
paths $1$ and $2$, respectively. Analogously, let the parenthesis vectors $\left(%
\begin{array}{c}
  1 \\
  0 \\
\end{array}%
\right)$ and $\left(%
\begin{array}{c}
  0 \\
  1 \\
\end{array}%
\right)$ represent the two \emph{polarization} states of a photon
polarized along a horizontal and a vertical direction,
respectively. Described in these terms, the $50/50$ beam splitter
performs a linear transformation restricted to the mode space
only; it can be represented by the $2 \times 2$ matrix
$\mathbf{B}$ as:
\begin{equation}\label{a100}
\mathbf{B} =  \frac{1}{\sqrt{2}} \left[%
\begin{array}{cc}
  1 & i \\
  i & 1 \\
\end{array}%
\right],
\end{equation}
where the relative phase shift of $\pi/2$ between the transmitted
and reflected amplitudes ensures unitarity:
$\mathbf{B}\mathbf{B}^\dagger = \mathbf{I}_2$. The attenuator A
 can be simply represented by a scalar function
$\exp(-\alpha)$, where $\alpha \geq 0$. The linear polarizer
$\mathrm{P}_H$ performs a linear transformation restricted to the
polarization space only. It can be represented by the projection
matrix $\mathbf{H}$ as:
\begin{equation}\label{a110}
\mathbf{H} = \left(%
\begin{array}{cc}
  1 & 0 \\
  0 & 0 \\
\end{array}%
\right).
\end{equation}
Finally, the polarization rotator R  can be represented by the
orthogonal matrix $\mathbf{R}$ in the polarization space as:
\begin{equation}\label{a120}
\mathbf{R} = \left. \left(%
\begin{array}{cc}
  \cos \theta & - \sin \theta \\
  \sin \theta & \cos \theta \\
\end{array}%
\right) \right|_{\theta = \frac{\pi}{2}} = \left(%
\begin{array}{cc}
  0 & - 1 \\
  1 & 0 \\
\end{array}%
\right) .
\end{equation}
If with $\mathbf{P}$ and $\mathbf{Q}$ we denote the two
complementary mode-space projectors
\begin{equation}\label{a130}
\mathbf{P} =  \left[%
\begin{array}{cc}
  1 & 0 \\
  0 & 0 \\
\end{array}%
\right], \qquad
\mathbf{Q} =  \left[%
\begin{array}{cc}
  0 & 0 \\
  0 & 1 \\
\end{array}%
\right],
\end{equation}
then the total $4 \times 4$ transmission matrix $\mathbf{T}$
representing $\mathcal{C}_\mathcal{M}$ can be written as:
\begin{equation}\label{a140}
\mathbf{T} =  \mathbf{R} \otimes \mathbf{P}.\mathbf{B} +
e^{-\alpha } \mathbf{H} \otimes \mathbf{Q}.\mathbf{B},
\end{equation}
where the low dot ``$\; . \;$'' denotes the ordinary matrix
product.

Let $| \mathrm{in} \rangle $ be the quantum state of a photon
entering $\mathcal{C}_\mathcal{M}$  through port 1:
\begin{equation}\label{a150}
| \mathrm{in} \rangle = \left(%
\begin{array}{c}
  \phi_H \\
  \phi_V \\
\end{array}%
\right) \otimes \left[%
\begin{array}{c}
  1 \\
  0 \\
\end{array}%
\right] \equiv | \phi \rangle \otimes |\psi \rangle,
\end{equation}
where $|\phi_H|^2 + |\phi_V|^2 =1$. In these terms, the two-mode
output state $|\mathrm{out}\rangle$ leaving
$\mathcal{C}_\mathcal{M}$ can be written as
\begin{equation}\label{a160}
\begin{array}{rcl}
 \displaystyle{|\mathrm{out}\rangle} & =& \displaystyle{\mathbf{T}
 |\mathrm{in}\rangle}\\\\
& =& \displaystyle{\mathbf{R}| \phi \rangle  \otimes
\mathbf{P}.\mathbf{B}|\psi \rangle + e^{-\alpha } \mathbf{H}| \phi
\rangle  \otimes \mathbf{Q}.\mathbf{B}|\psi \rangle
}\\\\
 & = & \displaystyle{ \frac{1}{\sqrt{2}} \left\{
  \left(%
\begin{array}{c}
  -\phi_V \\
  \phi_H \\
\end{array}%
\right) \otimes  \left[%
\begin{array}{c}
  1 \\
  0 \\
\end{array}%
\right] + i e^{ - \alpha}
 \left(%
\begin{array}{c}
  \phi_H \\
  0 \\
\end{array}%
\right) \otimes  \left[%
\begin{array}{c}
  0 \\
  1 \\
\end{array}%
\right] \right\}.
  }
\end{array}
\end{equation}
An elementary calculation shows that
\begin{equation}\label{a170}
\begin{array}{rcl}
 \displaystyle{|\mathrm{out}\rangle \langle \mathrm{out} |} & =& \displaystyle{\mathbf{T}
 |\mathrm{in}\rangle \langle \mathrm{in} | \mathbf{T}^\dagger}\\\\
& =& \displaystyle{\mathbf{R}| \phi \rangle \langle \phi |
\mathbf{R}^\dagger
 \otimes
\mathbf{P}.\mathbf{B}|\psi \rangle \langle \psi|\mathbf{B}^\dagger  .\mathbf{P}^\dagger}\\\\
& & \displaystyle{+ e^{-2 \alpha } \mathbf{H}| \phi \rangle
\langle \phi | \mathbf{H}^\dagger
 \otimes
\mathbf{Q}.\mathbf{B}|\psi \rangle \langle \psi|\mathbf{B}^\dagger  .\mathbf{Q}^\dagger}\\\\
& & \displaystyle{+ \bigl\{ e^{- \alpha } \mathbf{R}| \phi \rangle
\langle \phi | \mathbf{H}^\dagger
 \otimes
\mathbf{P}.\mathbf{B}|\psi \rangle \langle \psi|\mathbf{B}^\dagger
.\mathbf{Q}^\dagger \bigr.}\\\\
& & \displaystyle{ \; \quad \bigl.+ \mathrm{H.c.} \bigr\}},
\end{array}
\end{equation}
where H.c. stands for Hermitian conjugate.

Since, by hypothesis,  detector $\mathrm{D}_a$ does not
distinguish a photon exiting port $\mathbf{1}$ from a photon
exiting port $\mathbf{2}$, the \emph{detected} output state can be
obtained from $|\mathrm{out}\rangle \langle \mathrm{out} |$ by
tracing over the detected but unresolved position states:
\begin{equation}\label{a180}
 \mathrm{Tr} \bigl[ |\mathrm{out}\rangle \langle \mathrm{out} |
 \bigr] = \frac{1}{2}\left( \mathbf{R}| \phi \rangle \langle
\phi | \mathbf{R}^\dagger + e^{-2 \alpha } \mathbf{H}| \phi
\rangle \langle \phi | \mathbf{H}^\dagger \right),
\end{equation}
where trivially $\mathrm{Tr} \bigl[ \mathbf{P}.\mathbf{B}|\psi
\rangle \langle \psi|\mathbf{B}^\dagger .\mathbf{Q}^\dagger \bigr]
= 0$, and $\mathrm{Tr} \bigl[ \mathbf{P}.\mathbf{B}|\psi \rangle
\langle \psi|\mathbf{B}^\dagger .\mathbf{P}^\dagger \bigr] = 1/2 =
\mathrm{Tr} \bigl[ \mathbf{Q}.\mathbf{B}|\psi \rangle \langle \psi
|\mathbf{B}^\dagger .\mathbf{Q}^\dagger \bigr] $.
If we define $J_\mathrm{out} \equiv  \mathrm{Tr} \bigl[
|\mathrm{out}\rangle \langle \mathrm{out} | \bigr]$ and
$\rho_{\mathrm{in}} \equiv |\phi \rangle \langle \phi|$, then we
can rewrite Eq. (\ref{a180}) as
\begin{equation}\label{a190}
\begin{array}{rcl}
J_\mathrm{out} & = & \displaystyle{\frac{1}{2}\left(
\mathbf{R}\rho_{\mathrm{in}} \mathbf{R}^\dagger + e^{-2 \alpha }
\mathbf{H}\rho_{\mathrm{in}} \mathbf{H}^\dagger \right)}\\\\
& = & \displaystyle{\frac{1}{2 p}\left( \mathbf{M}_2
\rho_{\mathrm{in}} \mathbf{M}_2^\dagger + \frac{p \, e^{-2 \alpha
}}{2(1-p)} \mathbf{M}_3 \rho_{\mathrm{in}} \mathbf{M}_3^\dagger
\right)},
\end{array}
\end{equation}
where Eqs. (\ref{a20}),  (\ref{a110}), and (\ref{a120}) have been
used. If we choose $\alpha = \alpha(p)$ such that
\begin{equation}\label{a200}
 \frac{p \, e^{-2
\alpha }}{2(1-p)} =1  \quad \Rightarrow  \quad \alpha(p) =
-\frac{1}{2} \ln \frac{2(1-p)}{p},
\end{equation}
then Eq. (\ref{a190}) can be rewritten as
\begin{equation}\label{a210}
J_\mathrm{out} = \frac{1}{2p} \sum_{\mu = 0}^3 \mathbf{M}_\mu
\rho_\mathrm{in} \mathbf{M}_\mu^\dagger = \frac{1}{2p}
\rho_\mathrm{out},
\end{equation}
where Eq. (\ref{a10}) has been used. Note that $\alpha(p) \geq 0
\quad$ for $2/3 \leq p \leq 1$, as expected for an attenuator.
Equation (\ref{a210}) shows that the scheme shown in Fig.
\ref{fig:B} actually implements the map $\mathcal{M}$. Moreover,
from Eqs. (\ref{a150},\ref{a190}-\ref{a200}) it follows that
\begin{equation}\label{a220}
\begin{array}{rcl}
\displaystyle{\mathrm{Tr} \left( J_\mathrm{out} \right)}
& = & \displaystyle{ \frac{1}{2} \left[ 1 + \frac{2(1-p)}{p}
|\phi_H|^2 \right]},\\\\
& \therefore &  \displaystyle{ \frac{1}{2} \leq\mathrm{Tr} \left(
J_\mathrm{out} \right) \leq 1},
\end{array}
\end{equation}
for $2/3 \leq p \leq 1$ and $0 \leq |\phi_H| \leq 1$. This means
that even in the worst case ($p=1$) there is still a $50 \%$ of
probability to detect a photon in our scheme.
\section{Rigorous QED  treatment}
It was pointed out \cite{Ziman2} that the map $\mathcal{M}$
corresponds to a \emph{non-physical} quantum channel
$\mathcal{C}_\mathcal{M}$. Conversely, in the previous section we
have shown that a \emph{physical} linear optical implementation of
$\mathcal{C}_\mathcal{M}$ is actually feasible. The resolution of
this apparent paradox lies in the conceptual difference that
exists between the ``quantum state of two qubits'', and the
``\emph{measured} quantum state of two qubits''. The latter can be
reconstructed only after Alice and Bob have performed coincidence
measurements \cite{Abou}, that is only \emph{after} they have
established a communication and have compared their own
experimental results. Therefore, a \emph{measured} MEMS state
generated by a local channel does not raise any causality issue.
In this spirit,  we will soon show how the  map $\mathcal{M}$ can
be derived from an all-unitary model for the channel
$\mathcal{C}_\mathcal{M}$ (Fig. 3). Such a unitary channel reduces
to a non-unitary one when Alice restricts her measurements to two
output ports only ($\mathbf{1}$ and $\mathbf{2}$), leaving the
other two ($\mathbf{3}$ and $\mathbf{4}$) undetected. However,
note that in principle Alice could use an additional detector
$\mathrm{D}_{a}^{(\mathbf{34})}$ coupled to ports $\mathbf{3}$ and
$\mathbf{4}$ to generate a ``conditional''  MEMS state: When a
photon pair is created by the down-converter and detector
$\mathrm{D}_{a}^{(\mathbf{34})}$  does not fire, then a
conditional MEMS state is being transmitted through the channel.

 Let indicate with ${a}_{i \alpha}$ and
${b}_{\alpha}$ the annihilation operators of photons $a$ and $b$,
respectively. Greek indexes $\alpha, \beta, \ldots \in \{ 0,1\}$
label \emph{polarization} modes of the field, while Latin indexes
$ i,j, \ldots \in \{ 1, 2,3 ,4\}$ label \emph{spatial} modes of
the field. The latter modes represent the four paths shown in Fig.
3.
\begin{figure}[!hr]
\includegraphics[angle=0,width=8truecm]{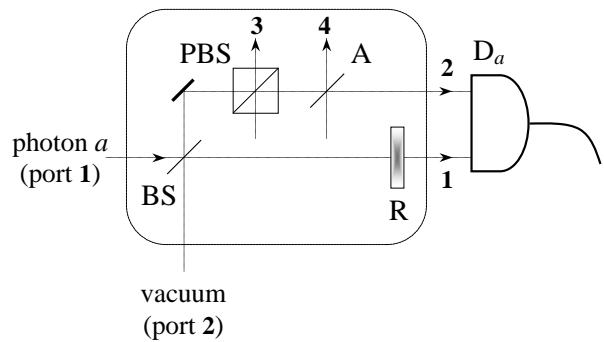}
\caption{\label{fig:3} All-unitary linear optical realization of
the quantum channel $\mathcal{C}_\mathcal{M}$. PBS is a polarizing
beam splitter that transmits only horizontally polarized photons.
The beam attenuator A is implemented by a variable-reflectivity
beam splitter. The two additional modes $\mathbf{3}$ and
$\mathbf{4}$ ensure the whole unitary nature of
$\mathcal{C}_\mathcal{M}$.}
\end{figure}
Described in these terms, the two-photon input singlet state can
be written as
\begin{equation}\label{a280}
| \mathrm{in} \rangle = \frac{1}{\sqrt{2}} \left( {a}_{ {1}
0}^\dagger {b}_{1}^\dagger - {a}_{ {1} 1}^\dagger {b}_{0}^\dagger
\right)|0\rangle,
\end{equation}
where $|0\rangle$ denotes the vacuum state. Each linear optical
element present in the quantum channel shown in Fig. 3, can be
represented by a unitary operator ${U}$ that evolves the state
vector $| \psi \rangle$ as
\begin{equation}\label{a290}
| \psi \rangle \mapsto U| \psi \rangle,
\end{equation}
and the operator $X$ either as
\begin{equation}\label{a300}
X \mapsto U^\dagger X U,
\end{equation}
or as
\begin{equation}\label{a305}
X \rightarrowtail U X U^\dagger.
\end{equation}
  In particular, if the annihilation operator  ${a}_{i
\alpha}$ evolves as
\begin{equation}\label{a310}
{a}_{i \alpha} \mapsto U^\dagger {a}_{i \alpha} U = \sum_{
{j}=1}^4 \sum_{\beta=0}^1 S_{ {i} \alpha, { j} \beta} {a}_{j
\beta},
\end{equation}
then it is easy to see that
\begin{equation}\label{a320}
{a}_{i \alpha}^\dagger \rightarrowtail U {a}_{i \alpha}^\dagger
U^\dagger = \sum_{ {j}=1}^4 \sum_{\beta=0}^1 S_{  j \beta,  i
\alpha} {a}_{j \beta}^\dagger.
\end{equation}
Within this formalism, the  beam splitter BS  is  described by the
matrix
\begin{equation}\label{a330}
S_{  j \beta,  i \alpha} = \mathbf{B}_{ji}\delta_{\beta \alpha}
\end{equation}
where   $\mathbf{B}$ is explicitly given in Eq. (\ref{a100}). This
leads to the field operators transformation
\begin{equation}\label{a340}
{a}_{1 \alpha}^\dagger \rightarrowtail \frac{1}{\sqrt{2}} \left(
{a}_{1 \alpha}^\dagger + i {a}_{2 \alpha}^\dagger \right),
\end{equation}
that modifies the input state $| \mathrm{in} \rangle$ to:
\begin{equation}\label{a350}
| \mathrm{in} \rangle \mapsto  \frac{1}{2} \left( {a}_{1
0}^\dagger {b}_{1}^\dagger + i {a}_{2 0}^\dagger {b}_{1}^\dagger -
{a}_{ 1 1}^\dagger {b}_{0}^\dagger - i {a}_{2 1}^\dagger
{b}_{0}^\dagger \right)|0\rangle .
\end{equation}
The effect of  the rotator $\mathrm{R}$ is very simple:
\begin{equation}\label{a360}
{a}_{1 0}^\dagger \rightarrowtail {a}_{1 1}^\dagger, \qquad {a}_{1
1}^\dagger \rightarrowtail -{a}_{1 0}^\dagger,
\end{equation}
 and it changes the two-photon states to
\begin{equation}\label{a370}
| \mathrm{in} \rangle \mapsto  \frac{1}{2} \left( {a}_{1
1}^\dagger {b}_{1}^\dagger + i {a}_{2 0}^\dagger {b}_{1}^\dagger +
{a}_{ 1 0}^\dagger {b}_{0}^\dagger - i {a}_{2 1}^\dagger
{b}_{0}^\dagger \right)|0\rangle .
\end{equation}
Next, the polarizing beam splitter PBS can be described by a $4
\times 4$ unitary matrix that couples both spatial  and
polarization modes, as
\begin{equation}\label{a380}
\left(%
\begin{array}{c}
  {a}_{2 0}^\dagger \\
  {a}_{2 1}^\dagger \\
  {a}_{3 0}^\dagger \\
  {a}_{3 1}^\dagger \\
\end{array}%
\right) \rightarrowtail
\left(%
\begin{array}{cccc}
  1 & 0 & 0 & 0 \\
  0 & 0 & 0 & i \\
  0 & 0 & 1 & 0 \\
  0 & i & 0 & 0 \\
\end{array}%
\right)
\left(%
\begin{array}{c}
  {a}_{2 0}^\dagger \\
  {a}_{2 1}^\dagger \\
  {a}_{3 0}^\dagger \\
  {a}_{3 1}^\dagger \\
\end{array}%
\right)
\end{equation}
As a result of this transformation, the two photon state after the
PBS can be written as
\begin{equation}\label{a390}
| \mathrm{in} \rangle \mapsto  \frac{1}{2} \left( {a}_{1
1}^\dagger {b}_{1}^\dagger + i {a}_{2 0}^\dagger {b}_{1}^\dagger +
{a}_{ 1 0}^\dagger {b}_{0}^\dagger + {a}_{3 1}^\dagger
{b}_{0}^\dagger \right)|0\rangle .
\end{equation}
Finally, the attenuator A can be described in a unitary fashion by
modelling it as a variable-reflectivity beam splitter such that:
\begin{equation}\label{a400}
{a}_{2 \alpha}^\dagger \rightarrowtail T {a}_{2 \alpha}^\dagger +
i R {a}_{4 \alpha}^\dagger,
\end{equation}
where $0 \leq T \leq 1$, and $T^2 + R^2 =1$. This last optical
element produces the output state $ | \mathrm{out} \rangle$, where
\begin{equation}\label{a410}
\begin{array}{rcl}
 | \mathrm{out} \rangle & = & \displaystyle{ \frac{1}{2}  \left[ \left( {a}_{1 1}^\dagger
{b}_{1}^\dagger  + {a}_{ 1
0}^\dagger {b}_{0}^\dagger + i T {a}_{2 0}^\dagger {b}_{1}^\dagger \right) \right.} \\\\
&&  \left. + \left({a}_{3 1}^\dagger {b}_{0}^\dagger  - R  {a}_{4
0}^\dagger {b}_{1}^\dagger
\right) \right]|0\rangle \\\\
   & \equiv & | \psi_{12} \rangle + | \psi_{34} \rangle.
\end{array}
\end{equation}
In Eq. (\ref{a410})  $| \psi_{ij} \rangle$   denotes the
two-photon state restricted to the pair of  modes $(i,j)$, and
$\langle \psi_{12} | \psi_{34} \rangle =0$. Since each
transformation performed by each linear optical element present in
the quantum channel is unitary,  the output state $ | \mathrm{out}
\rangle$ is still normalized: $\langle \mathrm{out}|\mathrm{out}
\rangle =1$. Now we can trace over the spatial degrees of freedom
in the usual way obtaining:
\begin{equation}\label{a420}
\rho = \mathrm{Tr} \left[ | \mathrm{out} \rangle\langle
\mathrm{out} | \right] \equiv \rho_{12} + \rho_{34},
\end{equation}
where
\begin{equation}\label{a430}
\rho_{12}  = \frac{1}{4}\left(%
\begin{array}{cccc}
  1 & 0 & 0 & 1 \\
  0 & T^2 & 0 & 0 \\
  0 & 0 & 0 & 0 \\
  1 & 0 & 0 & 1 \\
\end{array}%
\right),
\end{equation}
and
\begin{equation}\label{a440}
\rho_{34}  = \frac{1}{4}\left(%
\begin{array}{cccc}
  0 & 0 & 0 & 0 \\
  0 & 1 - T^2 & 0 & 0 \\
  0 & 0 & 1 & 0 \\
  0 & 0 & 0 & 0 \\
\end{array}%
\right).
\end{equation}
 The \emph{total} density matrix $\rho$ has
trace equal to one, while the \emph{truncated} density matrices
$\rho_{12}$ and $\rho_{34}$ have nonunit trace:
\begin{subequations}
\begin{equation}\label{a450}
\mathrm{Tr} \left( \rho_{12}\right) = \frac{1}{2} \left( 1 +
T^2/2\right) \equiv \frac{1}{2p};
\end{equation}
\begin{equation}\label{a460}
\mathrm{Tr} \left( \rho_{34} \right) = \frac{1}{2} \left( 1-
{T^2}/{2} \right) \equiv 1 - \frac{1}{2p}.
\end{equation}
\end{subequations}
This simple result shows that each truncated density matrix
\emph{cannot} be generated by a trace-preserving map. In
particular, we easily recover our result Eq. (\ref{a90}) by
dividing Eq. (\ref{a430}) by   Eq. (\ref{a450}). This is the goal
of the present section.

 To conclude, it may be
instructive to calculate separately the \emph{reduced} density
matrices $ \left. \mathrm{Tr} \rho_{ij} \right|_f$, obtained by
tracing over the degrees of freedom of photon $f$, where $( i,j)
\in \{ (1,2),(3,4)\}$ and $f = a,b$:
\begin{equation}\label{a470}
\begin{array}{ll}
\left. \mathrm{Tr} \rho_{12} \right|_a = \displaystyle{\frac{1}{4} }\left(%
\begin{array}{cc}
  1 & 0 \\
  0 & 1 + T^2 \\
\end{array}%
\right), & \left. \mathrm{Tr} \rho_{12} \right|_b = \displaystyle{\frac{1}{4} } \left(%
\begin{array}{cc}
  1+T^2 & 0 \\
  0 & 1  \\
\end{array}%
\right), \\\\
\left. \mathrm{Tr} \rho_{34} \right|_a = \displaystyle{\frac{1}{4} } \left(%
\begin{array}{cc}
  1 & 0 \\
  0 & 1 - T^2 \\
\end{array}%
\right), & \left. \mathrm{Tr} \rho_{34} \right|_b = \displaystyle{\frac{1}{4} } \left(%
\begin{array}{cc}
  1 - T^2 & 0 \\
  0 & 1  \\
\end{array}%
\right). \\
\end{array}
\end{equation}
From these results we learn that
  \begin{subequations}
\begin{equation}\label{a480}
\left. \mathrm{Tr} \rho_{12} \right|_a  + \left. \mathrm{Tr} \rho_{34} \right|_a
 =  \displaystyle{\frac{1}{2} }\left(%
\begin{array}{cc}
  1 & 0 \\
  0 & 1  \\
\end{array}%
\right) = \left. \mathrm{Tr} \rho_\mathrm{s} \right|_a,
\end{equation}
\begin{equation}\label{a490}
\left. \mathrm{Tr} \rho_{12} \right|_b  + \left. \mathrm{Tr}
\rho_{34} \right|_b
 =  \frac{1}{2} \left(%
\begin{array}{cc}
  1 & 0 \\
  0 & 1  \\
\end{array}%
\right) = \left. \mathrm{Tr} \rho_\mathrm{s} \right|_b,
\end{equation}
\end{subequations}
that is, when \emph{all} spatial modes of the two photons are
properly accounted for, locality requirements are fully satisfied.
\section{Conclusions}
Equations (\ref{a70}), (\ref{a90}),  (\ref{a260}), (\ref{a270}),
and (\ref{a210}), are the main results of our preliminary work
 on generation and measurement of maximally
entangled mixed states. In this paper we have shown how it is
possible to generate both MEMS I and II two-qubit states from the
singlet state by using only \emph{local}, non-trace-preserving
quantum channels. Moreover, we provided for the scheme of a simple
linear optical experimental setup for the generation of photonic
MEMS I states. Such a scheme, which exploit spatial degrees of
freedom of the photons to induce decoherence, is currently being
tested in our laboratory.
\begin{acknowledgments}
We acknowledge Vladimir Bu\v{z}ek  for useful comments on the
manuscript. This project is supported by FOM.
\end{acknowledgments}
%
%
%

%\bibliography{sca}
%\newpage

%\newpage

\end{document}